# Effective approach for taking into account interactions of quasiparticles from the low-temperature behavior of a deformed fermion-gas model

**Abdullah Algin[1,2,*] and Ali Serdar Arikan[3]**

[1]Department of Physics, Eskisehir Osmangazi University, Meselik, 26480-Eskisehir, Turkey

[2]Centre for Quantum Research and Applications, Eskisehir Osmangazi University, Meselik, 26480-Eskisehir, Turkey

[3]Department of Physics, Sakarya University, Esentepe, 54187-Sakarya, Turkey

**Abstract**

A deformed fermion gas model aimed at taking into account thermal and electronic properties of quasiparticle systems is devised. The model is constructed by the fermionic Fibonacci oscillators whose spectrum is given by a generalized Fibonacci sequence. We first introduce some new properties concerning the Fibonacci calculus. We then investigate the low-temperature thermostatistical properties of the model, and derive many of the deformed thermostatistical functions such as the chemical potential and the entropy in terms of the model deformation parameters $p$ and $q$. We specifically focus on the $p,q$-deformed Sommerfeld parameter for the heat capacity of the model, and its behavior is compared with those of both the free-electron Fermi theory and the experimental data for some materials. The results obtained in this study reveal that the present deformed fermion model leads to an effective approach accounting for interaction and compositeness of quasiparticles, which have remarkable implications in many technological applications such as in nanomaterials.

## 1. Introduction

Studies on complex systems are often required to consider nonlinear quantum behavior of the system containing either bosons or fermions associated with their both compositeness and interactions. In this context, one of the efficient and effective ways to approach such quantum properties is to employ an appropriate deformed particle system with one or two deformation parameters [1-10]. It should be emphasized that such deformed particle systems have been also used to understand the nature of both interparticle interactions [11-16] and entanglement characteristics in composite particle (or quasiparticle) systems [17,18].

Although thermostatistical properties of gases of deformed fermions and deformed bosons have been extensively investigated in the literature [19-49], the recent developments specifically carried out by the two-parameter deformed oscillator systems have been emerged interesting consequences in order to explore possible implications of these systems within other concrete physical applications. In this framework, we should mention some notable applications of such systems in condensed matter physics and quantum information science. For instance, the two-parameter deformed bosonic oscillators algebra, called bosonic Fibonacci oscillators, has been used to investigate some thermal and electrical properties of a solid [50,51] and possible consequences of introducing bosonic Fibonacci oscillators in the Landau diamagnetism problem have been also discussed [52]. These studies have led to new physical insights to approach some nonlinear behavior observed in the characteristics of crystalline structures such that rather than the use of one-parameter deformed systems [53-55], deformed oscillator systems with two distinct deformation parameters can play different roles to describe the change in thermal and electrical conductivities of metals. Even more interesting is the fact that each deformation parameter can be associated with different types of phenomena such as disorders and impurities [50,51]. Another interesting example to mention is that choosing different deformation structure functions has been yielded important roles in controlling the decoherence process [56], which has implications in quantum computation and quantum communication.

On the other hand, the recent progress on two-parameter deformed oscillator systems have been shown that the physical results obtained from a two-parameter deformed boson system are not the same with the ones by using a one-parameter deformed boson system [50-52,57-59]. In particular, deformed oscillator algebra with two independent deformation parameters could bring extra advantages when it applies to discuss some nonlinearities (or nonidealities) in the concrete phenomonological physical models. Thus, all the above motivations reveal concisely some reasons why we should consider the two-parameter deformed boson or fermion oscillator system within the different perspectives.

The present article is first aimed to study different quantum algebraic properties of a two-parameter deformed fermion system called fermionic Fibonacci oscillators, and secondly to investigate the low-temperature thermostatistical properties of a gas of these two-parameter deformed fermion oscillators. In this framework, we introduce some new properties concerning the Fibonacci calculus based on the present deformed fermion model. In the low-temperature limit, we also derive many of the deformed thermodynamical functions of the model such as the chemical potential and the entropy in terms of the real independent deformation parameters $p$ and $q$. Particular emphasize is given to an analysis on the electronic heat capacity constant called the $p,q$-deformed Sommerfeld parameter $\gamma^{(p,q)}$ along with its comparison by both the experimental data and the free-electron Fermi theory values on some substances chosen. Moreover, we should mention that the high-temperature thermostatistics of the same model has been discussed in [60] and here, we further study on this model to find new results about its low-temperature behavior. Besides, we aimed to search for possible effects of fermionic $p,q$-deformation on an application to the electronic properties of materials containing quasiparticle states such as in heavy fermion systems.

This article is organized as follows. In Section II, we present the fermionic Fibonacci oscillators model, and introduce some new properties regarding the Fibonacci calculus related to the model. In Section III, we investigate the low-temperature thermostatistical properties of a deformed gas of $p,q$-fermions, and obtain the deformed thermodynamic relations. The last section is devoted to a comprehensive discussion on our results and concluding remarks.

## 2. The fermionic Fibonacci oscillators' model

We now present the two-parameter deformed fermionic Fibonacci oscillators model. The model algebra is defined by the following deformed anti-commutation relations [61]:

$$c_i c_j = -\frac{q}{p} c_j c_i, \qquad i < j,$$

$$c_i c_j^* = -qp\, c_j^* c_i, \qquad i \neq j,$$

$$c_i^2 = 0,$$

$$c_1 c_1^* + p^2 c_1^* c_1 = p^{2\hat{N}}, \tag{1}$$

$$c_i c_i^* + q^2 c_i^* c_i = c_{i+1} c_{i+1}^* + p^2 c_{i+1}^* c_{i+1}, \qquad i = 1, 2, ..., d-1,$$

$$c_d c_d^* + q^2 c_d^* c_d = q^{2\hat{N}},$$

where $c_i$ and $c_i^*$ are the deformed fermionic annihilation and creation operators, $\hat{N}_i$ is the fermion number operator. $q$ and $p$ are the real positive independent deformation parameters. The total deformed fermion number operator for this system is

$$\sum_{i=1}^{d} c_i^* c_i = [\hat{N}_1 + \hat{N}_2 + ... + \hat{N}_d] = [\hat{N}], \qquad (2)$$

whose spectrum is given by the generalized Fibonacci basic integers as

$$[n] \equiv [n]_{p,q} = \frac{(p^{2n} - q^{2n})}{(p^2 - q^2)}. \qquad (3)$$

But, each $\hat{N}_i$ in Eq. (2) can only have eigenvalues of 0 and 1. The representation structure for the elements of fermionic Fibonacci oscillator algebra is also discussed in [61]. This $p,q$-deformed fermion algebra reduces to the usual fermion algebra in the limit $p = q = 1$. The fermionic Fibonacci oscillators offer a unification of quantum oscillators related to quantum group fermionic field algebras, and it constitutes the most general quantum group invariant fermionic algebra with the symmetry group $SU_{p/q}(d)$. In [61], it was also proved that as far as quantum group invariance is concerned, the maximum number of deformation parameters should be just two.

For the fermionic Fibonacci oscillator algebra in Eqs. (1)-(3), the transformation from Fock observables to the configuration space can be done by the following elements of Fibonacci calculus:

$$c^* \Rightarrow x, \qquad c \Rightarrow \hat{D}_x^{(p,q)}, \qquad (4)$$

where $\hat{D}_x^{(p,q)}$ is the modified Fibonacci difference operator [58] for the model algebra in Eqs. (1)-(3) defined as

$$\hat{D}_x^{(p,q)} f(x) = \left\{ \frac{(q^2 - p^2)}{[\ln(q^2 / p^2)]} \right\} \partial_x^{(p,q)} f(x), \tag{5}$$

where $\partial_x^{(p,q)} f(x)$ is

$$\partial_x^{(p,q)} f(x) = \frac{f(q^2 x) - f(p^2 x)}{(q^2 - p^2) x}, \tag{6}$$

for an analytic function $f(x)$. Thus, the operator $\hat{D}_x^{(p,q)}$ can be regarded as a two-parameter extension of the Jackson derivative (JD) operator [62]. Now, we continue to further study Eq. (5), and introduce some new formulations in the framework of Fibonacci calculus. The action of the modified Fibonacci difference operator $\hat{D}_x^{(p,q)}$ in Eq. (5) on a monomial $f(x) = \alpha x^n$, where $n \geq 0$ and $\alpha$ is a real constant, can be derived as

$$\hat{D}_x^{(p,q)} (\alpha x^n) = \left\{ \frac{(q^2 - p^2)}{[\ln(q^2 / p^2)]} \right\} \alpha [n]_{p,q} x^{n-1}, \tag{7}$$

where the Fibonacci basic number $[n]_{p,q}$ is given in Eq. (3). Acting on the product of the functions $f(x) = \alpha x^n$ and $g(x) = \beta x^m$, where $(n, m) \geq 0$ and $(\alpha, \beta)$ are real constants, via the $p,q$-deformed derivative operator $\hat{D}_x^{(p,q)}$ in Eq. (5) reads

$$\hat{D}_x^{(p,q)} (\alpha x^n \beta x^m) = \left\{ \frac{(q^2 - p^2)}{[\ln(q^2 / p^2)]} \right\} \alpha \beta [n + m]_{p,q} x^{n+m-1}. \tag{8}$$

Morever, the $l$th power of $p,q$-derivative on a monomial $(\alpha x^n)$ gives

$$(\hat{D}_x^{(p,q)})^l (\alpha x^n) = \left\{ \frac{(q^2 - p^2)}{[\ln(q^2 / p^2)]} \right\} \alpha \frac{[n]_{p,q}!}{[n - l]_{p,q}!} x^{n-l}, \tag{9}$$

where $[n]_{p,q}! = [n]_{p,q} [n-1]_{p,q} [n-2]_{p,q} \ldots 1$, via Eq. (3). The $p,q$-derivative satisfies the following property:

$$\hat{D}_{\alpha x}^{(p,q)} f(x) = \frac{1}{\alpha} \hat{D}_x^{(p,q)} f(x) . \tag{10}$$

The $p,q$-deformed analog of the Leibnitz rule for the modified Fibonacci difference operator $\hat{D}_x^{(p,q)}$ in Eq. (5) is proved as

$$\hat{D}_x^{(p,q)}[f(x)g(x)] = g(p^2 x)[\hat{D}_x^{(p,q)} f(x)] + f(q^2 x)[\hat{D}_x^{(p,q)} g(x)]$$
$$= g(q^2 x)[\hat{D}_x^{(p,q)} f(x)] + f(p^2 x)[\hat{D}_x^{(p,q)} g(x)], \tag{11}$$

and it also satisfies

$$\hat{D}_x^{(p,q)}\left(\frac{f(x)}{g(x)}\right) = \frac{g(p^2 x)[\hat{D}_x^{(p,q)} f(x)] - f(p^2 x)[\hat{D}_x^{(p,q)} g(x)]}{g(q^2 x) g(p^2 x)}$$
$$= \frac{g(q^2 x)[\hat{D}_x^{(p,q)} f(x)] - f(q^2 x)[\hat{D}_x^{(p,q)} g(x)]}{g(q^2 x) g(p^2 x)} . \tag{12}$$

By making use of the Fibonacci basic number $[n]_{p,q}$ and its factorial $[n]_{p,q}!$, we now introduce the $p,q$-deformed exponential function $\exp_{p,q}(x)$ as

$$e_{p,q}(x) \equiv \exp_{p,q}(x) = \sum_{n=0}^{\infty} \frac{x^n}{[n]_{p,q}!} , \tag{13}$$

where the $p,q$-factorial $[n]_{p,q}!$ is described above. Also, the $p,q$-deformed logarithm function $\ln_{p,q}(x)$ is proved as

$$\ln_{p,q}(x) = -\sum_{n=1}^{\infty} \frac{(1-x)^n}{[n]_{p,q}} , \tag{14}$$

where $[n]_{p,q}$ is defined in Eq. (3). The *p,q*-deformed exponential function satisfies the relation

$$\hat{D}_x^{(p,q)} e_{p,q}(\alpha x) = \alpha\, e_{p,q}(\alpha x). \tag{15}$$

The above introduced relations constitute different elements of the Fibonacci calculus, which also serves as the *p,q*-extension of Jackson *q*-calculus. They could play some roles for studying the low-temperature thermostatistics of a gas of fermionic Fibonacci oscillators. In the next section, we mainly intend to observe possible new consequences of introducing second deformation parameter in the low-temperature thermodynamical behavior of the present deformed fermion gas. Hence, we expect to see that how such consequences affect some thermal and electronic properties of a given material such as a metal or an alloy for low temperatures.

## 3. Low-temperature thermodynamics of the model

Now, we study the low-temperature behavior of a gas of fermionic Fibonacci oscillators, which is called as the *p,q*-fermion gas model. Such a *p,q*-fermion gas model can be described by the following fermionic Hamiltonian:

$$\hat{H}_{p,q}^F = \sum_k (\varepsilon_k - \mu)\hat{N}_k, \tag{16}$$

where $\mu$ is the chemical potential and $\varepsilon_k$ is the kinetic energy of a particle in the state *k* with the number operator $\hat{N}_k$. Although this fermionic Hamiltonian has a standard form, it implicitly includes the fermionic *p,q*-deformation via the deformed number operator given in Eq. (2). The mean value of the *p,q*-deformed occupation number $f_{k,p,q}$ for our model can be determined by the following relation [63]:

$$[f_{k,p,q}] = \frac{1}{Z} Tr(e^{-\beta \hat{H}_{p,q}^F}[\hat{N}_k]) \equiv \frac{1}{Z} Tr(e^{-\beta \hat{H}_{p,q}^F} c_i^* c_i), \tag{17}$$

where $\beta = 1/k_B T$, $k_B$ is the Boltzmann constant and $T$ is the temperature of the system. Also, $Z = Tr(e^{-\beta \hat{H}^F_{p,q}})$ is the fermionic grand partition function. Using the Fock space properties of the fermionic Fibonacci oscillators algebra in Eqs. (1)-(3) and applying the cyclic property of the trace [21], we find

$$f_{k,p,q} = \frac{1}{\left|\ln(q^2/p^2)\right|} \left|\ln\left(\frac{e^{\beta(\varepsilon_k - \mu)} + q^2}{e^{\beta(\varepsilon_k - \mu)} + p^2}\right)\right|, \tag{18}$$

where $p \neq q$ and $(p,q) \in R^+$. This may also be called as the *p,q*-deformed Fermi-Dirac (FD) distribution function. It provides a remarkable approximation for describing an intermediate-statistics quasifermion gas model. It reduces to the standard FD function in the limit $p = q = 1$ after applying this limiting case to the algebraic relations in Eqs. (1)-(3). Although an analysis on the high-temperature behavior of the model has been carried out by [60], here we continue to further study the same model in order to obtain different results by focusing on the low-temperature thermostatistical properties of the deformed gas of *p,q*-fermions.

To develop the low-temperature thermodynamics of such a two-parameter deformed fermion gas model, we consider the following relations for the *p,q*-deformed total number of particles $N^{(p,q)}(T)$ and the *p,q*-deformed total energy of the system $U^{(p,q)}(T)$:

$$N^{(p,q)}(T) = \int_0^\infty d\varepsilon \, g(\varepsilon) f(\varepsilon, T, p, q), \tag{19}$$

$$U^{(p,q)}(T) = \int_0^\infty \varepsilon \, d\varepsilon \, g(\varepsilon) f(\varepsilon, T, p, q), \tag{20}$$

where we have assumed the function of density of states $g(\varepsilon) = C\sqrt{\varepsilon}$, where $C = (V/2\pi^2)(2m/\hbar^2)^{3/2}$. Here the mass $m \equiv m(p,q)$ actually stands for the mass of the *p,q*-deformed quasifermionic particles in our model confined in a three-dimensional volume *V*. These integrals are of the form $K(p,q) = \int_0^\infty d\varepsilon \, h(\varepsilon) f(\varepsilon, T, p, q)$. Although, for low temperatures, one may consider the limiting case $z = \exp(\beta\mu) \gg 1$, where *z* is the fugacity, this situation is different for some applications in Bose systems. For instance, in a Bose-Einstein condensate, $z \leq 1$ corresponds to a low critical temperature. Such an observation is stressed here, since both the quasibosonic and the quasifermionic contributions to the total $p,q$-deformed quasiparticle heat capacity in the present model will be considered in the last section. At very low temperatures, the $p,q$-deformed integral $K(p,q)$ can be calculated by using the integration by parts as

$$K(p,q) = f(\varepsilon,T,p,q)H(\varepsilon)\Big|_0^\infty - \int_0^\infty H(\varepsilon)\frac{\partial f(\varepsilon,T,p,q)}{\partial \varepsilon}d\varepsilon, \qquad (21)$$

where the function $H(\varepsilon)$ is defined as $H(\varepsilon) = \int h(\varepsilon)d\varepsilon$. The first term on the right hand side of this equation vanishes, because of the fact that either $H(\varepsilon) = 0$ at $\varepsilon = 0$ or $f(\varepsilon,T,p,q) = 0$ in the limiting case $\varepsilon \to \infty$. So, we have

$$K(p,q) = -\int_0^\infty H(\varepsilon)\frac{\partial f(\varepsilon,T,p,q)}{\partial \varepsilon}d\varepsilon. \qquad (22)$$

On the other hand, the function $H(\varepsilon)$ can be expanded in Taylor series up to the first three terms, when this function does mainly vary in the vicinity of $\varepsilon = \mu$. This assumption gives

$$H(\varepsilon) = H(\mu) + (\varepsilon - \mu)\left(\frac{\partial H(\varepsilon)}{\partial \varepsilon}\right)\Bigg|_{(\varepsilon=\mu)} + \frac{1}{2}(\varepsilon - \mu)^2\left(\frac{\partial^2 H(\varepsilon)}{\partial \varepsilon^2}\right)\Bigg|_{(\varepsilon=\mu)}, \qquad (23)$$

which leads to the following expression from Eq. (22):

$$K(p,q) = -H(\mu)\int_0^\infty \frac{\partial f(\varepsilon,T,p,q)}{\partial \varepsilon}d\varepsilon - \left(\frac{\partial H(\varepsilon)}{\partial \varepsilon}\right)\Bigg|_{(\varepsilon=\mu)}\int_0^\infty (\varepsilon-\mu)\frac{\partial f(\varepsilon,T,p,q)}{\partial \varepsilon}d\varepsilon$$

$$-\frac{1}{2}\left(\frac{\partial^2 H(\varepsilon)}{\partial \varepsilon^2}\right)\Bigg|_{(\varepsilon=\mu)}\int_0^\infty (\varepsilon-\mu)^2\frac{\partial f(\varepsilon,T,p,q)}{\partial \varepsilon}d\varepsilon. \qquad (24)$$

In the low-temperature limit, using a change of variables in the $p,q$-deformed FD distribution function $f(\varepsilon,T,p,q)$ with $x=\beta(\varepsilon-\mu)$, $dx=\beta d\varepsilon$ and practically changing the lower bound of these integrals in Eq. (24) by $(-\infty)$, we obtain the result

$$K(p,q) = H(\mu) - \frac{1}{\beta}\left(\frac{\partial H(\varepsilon)}{\partial \varepsilon}\right)\bigg|_{(\varepsilon=\mu)} \int_{-\infty}^{+\infty} x \frac{\partial f(x,p,q)}{\partial x} dx - \frac{1}{2\beta^2}\left(\frac{\partial^2 H(\varepsilon)}{\partial \varepsilon^2}\right)\bigg|_{(\varepsilon=\mu)} \int_{-\infty}^{+\infty} x^2 \frac{\partial f(x,p,q)}{\partial x} dx. \quad (25)$$

If we define a $p,q$-deformed integral representation $I^{FFO}(p,q)$ for our model as $I^{FFO}(p,q) = \int_{-\infty}^{+\infty} x^2 (\partial f(x,p,q)/\partial x) dx$, where FFO stands for the system of fermionic Fibonacci oscillators, we then have

$$K(p,q) = H(\mu) - \frac{1}{\beta}\left(\frac{\partial H(\varepsilon)}{\partial \varepsilon}\right)\bigg|_{(\varepsilon=\mu)} \int_{-\infty}^{+\infty} x \frac{\partial f(x,p,q)}{\partial x} dx - \frac{1}{2\beta^2}\left(\frac{\partial^2 H(\varepsilon)}{\partial \varepsilon^2}\right)\bigg|_{(\varepsilon=\mu)} I^{FFO}(p,q). \quad (26)$$

For very low temperatures, $[\partial f(x,p,q)/\partial x]$ does mainly vary in the vicinity of $\varepsilon = \mu$. Therefore, the second term on the right hand side can be practically taken as zero, because the function under this integral is odd. Under these approximations, we find out

$$K(p,q) = H(\mu) - \frac{1}{2\beta^2}\left(\frac{\partial^2 H(\varepsilon)}{\partial \varepsilon^2}\right)\bigg|_{(\varepsilon=\mu)} I^{FFO}(p,q), \quad (27)$$

which can be rewritten by considering the definition of $H(\varepsilon)$ after Eq. (21) as

$$K(p,q) = \int_0^\mu h(\varepsilon) d\varepsilon - \frac{1}{2\beta^2}\left(\frac{\partial h(\varepsilon)}{\partial \varepsilon}\right)\bigg|_{(\varepsilon=\mu)} I^{FFO}(p,q), \quad (28)$$

where the values of $I^{FFO}(p,q)$ depend on the deformation parameters $p$ and $q$, and it has always negative values for the range $(p,q)<1$. Its values also change more rapidly after the values of deformation parameters $q=0.69$, $p=0.70$ in the interval $0<(p,q)<1$. Using Eq. (18), it can be rewritten as

$$I^{FFO}(p,q) = \int_{-\infty}^{+\infty} x^2 dx \left\{ \frac{1}{[\ln(q^2/p^2)]} \left[ \frac{1}{(1+q^2 e^{-x})} - \frac{1}{(1+p^2 e^{-x})} \right] \right\}, \tag{29}$$

where we have suppressed the absolute value signs for the sake of simplicity. It can be solved via some $p,q$-analog of the polylogarithm function such that it provides algebraically a consistent solution just for the case $(p,q)<1$. Due to this fact, we will consider the interval $0<(p,q)<1$ for the rest of calculations in this work.

We should also stress that the expression in Eq. (24) has some remarkable implications such that not only it reveals a $p,q$-deformed extension of the Sommerfeld expansion method related to the low-temperature thermostatistics of the model, but also generalizes the one-parameter deformed fermion gas results introduced recently in [37,38,55,64].

In order to further analyze the $p,q$-deformed thermostatistical characteristics of our model at low temperatures, we first evaluate the relation given in Eq. (19). By means of Eq. (28) and taking $h(\varepsilon) = g(\varepsilon)$, where $g(\varepsilon)$ is described above, we obtain the following expression for the $p,q$-deformed total number of particles for our model:

$$N^{(p,q)}(T) = \frac{V}{3\pi^2}\left(\frac{2m}{\hbar^2}\right)^{3/2} \mu^{3/2} \left[ 1 - \frac{3}{8} I^{FFO}(p,q)\left(\frac{k_B T}{\mu}\right)^2 \right], \tag{30}$$

which leads to the following result for the chemical potential in the zeroth approximation:

$$\mu(0,p,q) = \left(\frac{\hbar^2}{2m}\right)\left(\frac{3\pi^2 N^{(p,q)}(0)}{V}\right)^{2/3}, \tag{31}$$

which equals to the standard Fermi energy $\varepsilon_F$ for an undeformed fermion gas corresponding to the limit $p = q = 1$. This result implies that quantum deformation of fermions is just a finite temperature effect. In the first approximation, we also obtain

$$\mu(T,p,q) = \varepsilon_F \left[ 1 + \frac{1}{4} I^{FFO}(p,q)\left(\frac{k_B T}{\varepsilon_F}\right)^2 \right]. \tag{32}$$

For comparison, in Figs. 1 and 2, we show the plots of the $p,q$-deformed chemical potential $(\mu(T,p,q)/\varepsilon_F)$ and the chemical potential $(\mu(T,1,1)/\varepsilon_F)$ of an undeformed fermion gas as a function of $(k_B T/\varepsilon_F)$ for several values of the deformation parameters $p$ and $q$ for the cases $(p,q)<1$ and $p=q=1$, respectively. The values of the $p,q$-deformed chemical potential $(\mu(T,p,q)/\varepsilon_F)$ for the interval $0<(p,q)<1$ increase with the values of second deformation parameter $p$ at the same values of $(k_B T/\varepsilon_F)$ as shown in Fig. 1. Also, when compared with the case $p=q=1$ in Fig. 2, the values of this function are lower than those of an undeformed fermion gas at the same values of $(k_B T/\varepsilon_F)$.

Furthermore, from Eqs. (20), (28) and (32) together with $h(\varepsilon)=\varepsilon\, g(\varepsilon)$, the $p,q$-deformed total energy of the model in the low-temperature regime becomes

$$U^{(p,q)}(T) = \frac{3}{5} N^{(p,q)}(0)\,\varepsilon_F \left[1 - \frac{5}{4} I^{FFO}(p,q)\left(\frac{k_B T}{\varepsilon_F}\right)^2 \right]. \tag{33}$$

Hence, the $p,q$-deformed quasifermionic heat capacity of the model can be obtained from the thermodynamic relation $C^{(p,q)}_{\substack{quasi\\fermionic}}(T) = \left(\partial U^{(p,q)}(T)/\partial T\right)_{V,N}$, which leads to the result

$$C^{(p,q)}_{\substack{quasi\\fermionic}}(T) = -\frac{3}{2} k_B N^{(p,q)}(0) I^{FFO}(p,q)\left(\frac{k_B T}{\varepsilon_F}\right). \tag{34}$$

This actually represents a quasifermionic (or quasi-electronic) heat capacity relation for our model, since the fermionic Fibonacci oscillators algebra in Eqs. (1)-(3) constitutes originally a quasifermion algebra. The deformed quasifermionic heat capacity depends linearly on the temperature $T$ at sufficiently low temperatures and goes to zero in the limit $T=0$ satisfying the third law of thermodynamics. Moreover, the result in Eq. (34) can be rewritten as

$$C^{(p,q)}_{\substack{quasi\\fermionic}}(T) = \gamma^{(p,q)} T, \tag{35}$$

where the quasifermionic heat capacity constant $\gamma^{(p,q)}$, called the $p,q$-deformed Sommerfeld parameter, is

$$\gamma^{(p,q)} = \left[ -\frac{3}{2} \frac{k_B^2 N^{(p,q)}(0)}{\varepsilon_F} \right] I^{FFO}(p,q). \tag{36}$$

This constitutes one of our main results, and as far as we know from the literature, this is the first attempt to introduce such a quasi-electronic heat capacity constant for the system containing $p,q$-fermions. We believe that such a model with two deformation parameters can have different advantages to take into account some nonlinearities (or nonideality factors) observed in composite fermion materials. We should note that Eq. (36) reduces to the undeformed electronic heat capacity constant $\gamma^{(1,1)} = (\pi^2 N^{(1,1)}(0) k_B^2 / 2\varepsilon_F)$ in the limit $p = q = 1$ upon considering the discussion made after Eq. (18). In order to make a comparison with the results of both the free-electron theory and the experimental data for the materials chosen, we apply the result in Eq. (36) to some materials such as nickel (*Ni*) and cobalt (*Co*), which have remarkable applications in many areas of interest. In Figs. 3 and 4, we present the behavior of the $p,q$-deformed Sommerfeld parameter $\gamma^{(p,q)}$ of the materials *Ni* and *Co* as a function of the deformation parameters $p$ and $q$ for the range $(p,q) < 1$, respectively. The $p,q$-deformed Sommerfeld parameter for these materials are found as $\gamma^{(p,q)} = 7.02$ (*mJ/molK²*) for *Ni* with $p = 0.8934$, $q = 0.5768$ and $\gamma^{(p,q)} = 4.98$ (*mJ/molK²*) for *Co* with $p = 0.74$, $q = 0.65$, respectively. Note that the values of both the free-electron theory $\gamma^{(1,1)}$ and the experimental $\gamma^{(\exp)}$ for the electronic heat capacity constants of these materials are also given in the following table:

Table 1.

Besides, the entropy for the $p,q$-fermion gas model can be found from $S = (U - F)/T$, where the Helmholtz free energy $F$ satisfies the thermodynamic relation $F = \mu N - PV$ with the pressure $P = (2U/3V)$. Therefore, using Eqs. (30) and (33), we derive the $p,q$-deformed entropy function $[S^{(p,q)}(T)/k_B N^{(p,q)}(0)]$ of the model at low temperatures as

$$\frac{S^{(p,q)}(T)}{k_B N^{(p,q)}(0)} = -\frac{3}{2} I^{FFO}(p,q) \left( \frac{k_B T}{\varepsilon_F} \right). \tag{37}$$

For comparison, in Figs. 5 and 6, we show the plots of the *p,q*-deformed entropy function $[S^{(p,q)}(T)/k_B N^{(p,q)}(0)]$ and the entropy function $[S^{(1,1)}(T)/k_B N^{(1,1)}(0)]$ of an undeformed fermion gas as a function of $(k_B T/\varepsilon_F)$ for several values of the deformation parameters *p* and *q* for the cases $(p,q) < 1$ and $p = q = 1$, respectively. At the same values of $(k_B T/\varepsilon_F)$, the *p,q*-deformed entropy values of our model are larger than those of an undeformed fermion gas as shown in Figs. 5 and 6. We also observe from Fig. 5 that the *p,q*-deformed entropy values of the model decrease with the value of second deformation parameter *p* for the range $(p,q) < 1$. Such a result allows us to conclude that the larger deformation results in the lesser chaoticity in the system of deformed quasifermions.

Before closing this section, we should emphasize that the results obtained above show the effects of fermionic quantum deformation on the thermal and electronic properties of a quasifermion system at low temperatures. They also reveal notable differences between the results of the present deformed quasifermion model and the free-electron Fermi gas model [65-70]. Other concluding remarks and potential implications for the above results will be comprehensively discussed in the next section.

## 4. Discussion and conclusions

In this work, we investigated possible consequences of introducing fermionic quantum deformation by applying it into thermal and electronic properties of materials at low temperatures. In this context, we first studied the two-parameter extension of quantum calculus, called Fibonacci calculus, and have introduced new properties in Eqs. (7)-(15) concerning the Fibonacci calculus related to the present model. Second, starting with the two-parameter deformed FD statistics in Eq. (18), we derived many of the important thermostatistical functions of the present deformed quasifermion model in terms of the deformation parameters *p* and *q* in the low-temperature limit. Our results not only reveal the low-temperature behavior of a two-parameter deformed quasifermion gas model obeying intermediate-statistics but also provide interesting implications on the role of the parameters *p* and *q* for taking into account effective interactions of quasifermions and their compositeness. We should also stress that our findings exhibit many benefits of using the present deformed quasifermion model rather than the free-electron Fermi gas model to deal with the low-temperature thermodynamical properties of materials.

In the usual treatment of the free-electron Fermi gas, the heat capacity of metals is the sum of electron and phonon contributions [67]. Therefore, for our model, we can write the total *p,q*-deformed quasiparticle heat capacity as

$$C_{total}^{(p,q)}(T) = C_{\substack{quasi \\ fermionic}}^{(p,q)}(T) + C_{\substack{quasi \\ bosonic}}^{(p,q)}(T) = \gamma^{(p,q)} T + A^{(p,q)} T^3, \tag{38}$$

where the second term on the right hand side is a contribution from the quantization of lattice vibrations such as the *p,q*-deformed phonons for some deformed lattice model. Although, a discussion on the deformed lattice contribution to the total heat capacity of the system is beyond the scope of this work, we would like to mention a recent investigation on this subject. Indeed, using the *p,q*-deformed bosonic oscillators algebra, called bosonic Fibonacci oscillators [9], the thermal and electrical properties of a solid have been investigated in [50]. From an analysis on the (*p,q*)-deformed Debye solid through bosonic Fibonacci oscillators, they have found the following expression for the *p,q*-deformed phonon excitation term of the low-temperature heat capacity in a deformed Debye solid [50]:

$$C_{\substack{quasi \\ bosonic}}^{(p,q)}(T) = \frac{12\pi^4 k_B}{5} \left( \frac{T}{\Theta_{D_{p,q}}} \right)^3, \tag{39}$$

where $\Theta_{D_{p,q}}$ is the *p,q*-deformed Debye temperature. Finally, we can express the total *p,q*-deformed quasiparticle heat capacity for a material containing both fermionic and bosonic Fibonacci oscillators from Eqs. (38) and (39) as

$$C_{total}^{(p,q)}(T) = \gamma^{(p,q)} T + \frac{12\pi^4 k_B}{5} \left( \frac{T}{\Theta_{D_{p,q}}} \right)^3, \tag{40}$$

where the *p,q*-deformed Sommerfeld parameter $\gamma^{(p,q)}$ is defined in Eq. (36). Note that the first term in the right hand side of Eq. (40) arising from the contribution by quasifermions, i.e. quasi-electrons, is dominant rather than the second term for very low temperatures. In [50,51], the authors discussed possible effects of the bosonic *p,q*-deformation on the thermal and electrical conductivities of metals through bosonic Fibonacci oscillators, and found that these deformed bosonic oscillators may act as defects or impurities in the crystal lattice. In addition to this, the results obtained in a study of the (*p,q*)-deformed Debye solid [50,51] revealed that each deformation parameter can be associated to different types of deformations such as impurities and disorders. Even more interesting is the fact that as previous works reported the similar issues from different perspectives [15,29,30,36,58,59], the results of [50-52,60] with two deformation parameters differ from that of earlier ones by considering only one deformation parameter [19-28,31-35,37-49]. Hence, the present study on the fermionic *p,q*-deformation applying to the thermal and electronic properties of materials at low temperatures has been put forward new results to support the idea that inserting two deformation parameters in physical applications exposes different physical consequences. Moreover, we should emphasize that as is exemplified in Table 1, some other failures as well as discrepancies between the standard theories such as the free-electron Fermi gas and their related experiments can be overcome by using the present deformed quasifermion model obeying intermediate-statistics with two independent deformation parameters.

Since, we are focusing on the low-temperature thermostatistical properties of our model, we now wish to set up a link between the thermal effective mass of a deformed quasifermionic particle and the model deformation parameters. Also, we should note that as is shown in Table 1, for most of other materials such as transition metals and rare earth compounds, the calculated free-electron theory values of electronic heat capacity constant $\gamma^{(1,1)}$ do not agree with experimental values [65-67]. We can construct a link by using the *p,q*-deformed quasifermionic heat capacity of the model given in Eqs. (34)-(36) along with the use of Eq. (31) and $g(\varepsilon)$, which is defined after Eq. (20). The *p,q*-deformed quasifermionic heat capacity of our model in Eq. (34) is inversely proportional to the Fermi energy $\varepsilon_F$, which is also inversely proportional to the mass *m* as shown in Eq. (31). All these considerations show that the *p,q*-deformed Sommerfeld parameter $\gamma^{(p,q)}$ in Eq. (36) is directly proportional to the mass of a deformed quasifermionic particle with $m \equiv m(p,q)$, which is also defined after Eq. (20). Therefore, we have $\gamma^{(p,q)} \propto m$. Since it is common to practically express the ratio of the measured $\gamma^{(\exp)}$ to the free particle values of electronic heat capacity as a ratio of a thermal effective mass to the free particle mass, the following relation can be described for our model:

$$\frac{m^*(p,q)}{m} = \frac{\gamma^{(\exp)}}{\gamma^{(p,q)}} \;, \tag{41}$$

where $m^*(p,q)$ is the thermal effective mass of a deformed quasifermion incorporated with the effects of interactions in the model. Using Eq. (36), the relation in Eq. (41) can be rewritten as

$$\frac{m^*(p,q)}{m} = \frac{[\gamma^{(\exp)}/\gamma^{(1,1)}]}{[(-3\pi^{-2})I^{FFO}(p,q)]} \;, \tag{42}$$

where the undeformed Sommerfeld parameter $\gamma^{(1,1)}$ is described after Eq. (36) and $I^{FFO}(p,q)$ is given in Eq. (29). Since it is very practical to exploit the ratio $[\gamma^{(\exp)}/\gamma^{(1,1)}]$ for a given material under consideration, the relation in Eq. (42) provides a remarkable connection between the parameters of fermionic deformation and the thermal effective mass of a quasifermion. This also implies that the fermionic $p,q$-deformation is related to phenomena due to effective interparticle interactions inside the model. Such an interpretation is a consequence of the fact that the ratio $[m^*(p,q)/m]$ is a measure of the effects originating from some interaction with lattice potential on the deformed quasifermions and effective interactions among quasifermions along with their interactions with quasibosonic lattice phonons. Such interaction effects lead to a thermal effective mass $m^*(p,q)$ for the deformed quasifermions within the model. This allows us to deduce a modified dispersion relation for the deformed quasifermions in our model as

$$\varepsilon(\vec{k},p,q) = \frac{\hbar^2 k^2}{2m^*(p,q)} \;, \tag{43}$$

where the constraint $k^2 = k_x^2 + k_y^2 + k_z^2$ is considered. Therefore, all the information about the details of effective interparticle interactions is encoded via the model deformation parameters $p$ and $q$. From the results in Eqs. (18) and (42), we conclude that these model parameters not only manage the deformed quantum statistics of quasifermionic particles, but also control the strength of effective interactions among quasifermions. In effect, some dramatic deviations of the ratio $(m^*/m)$ between the free-electron theory and experiments observed in real fermion systems such as in heavy fermion materials can efficiently be regulated by the model parameters $p,q$ of fermionic deformation in some nontrivial way as shown in Eqs. (36) and (42). Moreover, from the relations given in Eqs. (16), (42) and (43), the present deformed quasifermion model can be interpreted in such a way that the $p,q$-deformed noninteracting system is incorporated into the effects of interactions among its quasifermionic particles. Such consideration coming out by the fermionic deformation also supports and provides a new evidence to substantiate a recent idea based on a bosonic $q$-oscillator in [12,13] that it is possible to interpret the bosonic deformation as a $q$-deformed noninteracting system describing a nondeformed interacting physical system. Consequently, from the results obtained above, our model have a concise interpretation as the system of interacting quasiparticles through a $p,q$-deformed free quasifermionic gas.

In the light of the result given in Eq. (36), we now wish to discuss a remarkable connection between the model parameters $p,q$ and some physical property related to compositeness of deformed quasifermions. As we know from the usual treatment of free-electron Fermi gas [71-74] that the heat capacity measurements serve as a means to obtain the electron population density at the Fermi surface by using the following expression of an undeformed Sommerfeld parameter [66]:

$$\gamma^{(1,1)} = (0.169) Z \left(\frac{r_s}{a_0}\right)^2 \times 10^{-4} \quad cal/mol\, K^2, \tag{44}$$

where $a_0$ is the Bohr radius defined as $a_0 = \hbar^2/me^2$ and $Z$ is the number of valence electrons. Also, $r_s$ is defined as the radius of a sphere, whose volume is equal to the volume per conduction electron obeying $r_s = (3/4\pi n)^{1/3}$, where $n = N/V$. The compositeness parameter $r_s$ can also be viewed as the effective range between any two conduction electrons. If we associate with the physical quantities given by Eqs. (36) and (44), we then infer the following formula of the $p,q$-deformed Sommerfeld parameter for our model:

$$\gamma^{(p,q)} = (-3\pi^{-2}) \left[ (0.169) Z \left(\frac{r_s}{a_0}\right)^2 \times 10^{-4} \right] I^{FFO}(p,q), \tag{45}$$

where $I^{FFO}(p,q)$ is defined in Eq. (29). Therefore, this relation can be effectively used to take into account the parameter $r_s$ for a given material with its known values of $Z$ and $a_0$. If we insert the experimental value $\gamma^{(\exp)}$ of such a material into Eq. (45) instead of using the calculated value $\gamma^{(p,q)}$ in Eq. (36), we can then be able to determine the compositeness parameter $r_s$ for each quasiparticle within the material under consideration.

Furthermore, we conclude from all the results obtained above that our model lead implicitly to the following functional form of the deformation parameters $p$ and $q$ in view of their relations to physical characteristics of a given quasifermion system :

$$p = p(T, m^*, r_s), \qquad q = q(T, m^*, r_s). \qquad (46)$$

Thus, our approach based on the low-temperature thermostatistics of $p,q$-deformed analog of Fermi gas model establish an effective way accounting for both interactions of quasiparticles and their compositeness. These aspects can also be considered as nonideality or nonlinearity factors in the system under consideration. Therefore, the low-temperature behavior of nonlinear physical systems can be dealed with the properties of the present deformed quasifermion gas model, which shows us how to incorporate into the changes occuring in both thermal effective masses of quasiparticles and their interactions via the fermionic deformation. Our results also reveal that the model deformation parameters $p$ and $q$ serve as extra phenomenological means to control some discrepancies between theory and experiments, which can be useful for applications in the effective modeling of nonlinear behavior of interacting composite fermion (or quasifermion) systems such as observed in nanomaterials.

Apart from the present application of fermionic $p,q$-deformation to thermal and electronic properties of materials for low temperatures, we now would like to discuss other potential application areas of the model studied here. Since our model revealed the possibility of obtaining the details about effective quasifermion interactions of quantum statistical origin and some compositeness effects of deformed quasifermions, one could apply the results in Eqs. (18)-(46) to simulate the behavior of some exotic fermionic quasiparticle states such as the polaron (see e.g., [75]). One further direction to such a study could even be an integration of the bosonic Fibonacci oscillators, which can be associated to impurities in a crystal lattice [50,51]. Thus, interaction mechanism behind the special combinations of electrons and phonons, often referred to as polarons, could be understood better by modeling such type of quasiparticle systems with deformed quasibosons and deformed quasifermions.

Another potential application of the present deformed quasifermion model could be in the Kondo effect [76]. As we know that the Kondo effect involves an unusual scattering process of conduction electrons in a metal due to magnetic impurities, thereby increasing the resistance. As is exemplified in Table 1, there are large gaps between the free-electron theory value $\gamma^{(1,1)}$ and the experimental value $\gamma^{(exp)}$ for some other substances such as in heavy fermion materials. Hence, for very low temperatures and by viewing electrons as deformed quasifermions, our model can be used for a better understanding the dynamics behind the strong coupling regime occuring between localized magnetic impurities and quasi-electrons, which can have enormous values of thermal effective masses due to the interactions. Our results can also be applied to effectively interacting systems on other Kondo effect applications such as in a quantum dot.

One other potential application of the present deformed quasifermion model could be in the framework of a very recent development [77] on the new light with half-integral spin. Parthasarathy and Viswanathan [78,79] proposed that this newly discovered "light (photons) with half-integral spin" can be interpreted as $q$-deformed fermions without the exclusion principle. As we have recently discussed the high-temperature thermostatistical properties of the present deformed quasifermion gas model for both two and three spatial dimensions in [60], we should remark that both the results obtained in this study and the high-temperature properties introduced in [60] could be useful for further researches on understanding of the high and low temperature behavior of new photons with half-integer spin. Such exotic quasiparticle states could have an intermediate-statistics behavior to be discussed within the framework of a deformed gas of $p,q$-photons, whose underlying quantum thermodynamics can be modeled by the present approach.

Furthermore, one possible application of the present deformed quasifermion gas model can be in the field of cosmology. Since, our analysis develops an effective and efficient model endowed with both an intermediate-statistics behavior and two-parameter deformed thermostatistics formalism for low temperatures via the Fibonacci calculus. In 1993, Strominger proposed in [80] that the quantum statistics of charged, extremal black holes is a deformed one obeying a special $q$-deformed quantum commutation relation. Starting from such an idea, solution of the one and two-parameter deformed Einstein equations and its application to the extremal quantum black holes have been discussed in [81,82]. Hence, in the framework of such studies, our results in this work could lead to new physical insights in studies on a possible formulation of the quantum black hole thermodynamics, particularly for approximating the quantum statistics of horizon of black holes with an appropriate form for the intermediate-statistics similar to the one introduced in Eq. (18).

According to the above analysis, the model deformation parameters $p$ and $q$ can even play different roles such that one parameter can responsible for thermal effective masses of quasifermions implying a control on interparticle interactions, while the other one can affect some compositeness property such as the parameter $r_s$ implying a modification on the quasifermion population density in the system under consideration. The information about interaction and compositeness structure can even be simultaneously encoded in the model deformation parameters $p$ and $q$ with $p \neq 1$, $q \neq 1$ and $p \neq q$. As we have recently investigated the high temperature thermostatistical behavior of the same model containing its equation of state as a ($p,q$)-deformed virial expansion in both two and three spatial dimensions [60], both the present analysis and the results of [60] do also support the idea such that $p,q$-deformation of fermions leads to the system of interacting gas of quasi (or composite) particles. This can be interpreted in a way that $p,q$-deformation is related to both the effective interparticle interactions and compositeness of particles as long as we concerned with the deformed quantum statistics in a given quantum mechanical system. Hence, our results show considerable differences from both the one-parameter deformed fermion gas studies [12,13,17,19-28,32-35,37-43,53-55] and the free-electron Fermi theory [66,67]. Such differences reveal extra advantageous of the model, which also refers to a greater flexibility on the wide range of its deformation parameter spectrums in realistic phenomenological applications. These spectrums containing the values of $p,q$ in the model are guaranteed by the existence of more degrees of freedom in Eqs. (1)-(3), thereby increasing the adjustment range, i.e., it is possible to have rather different combinations for values of $p,q$ in order to calibrate one or more factors of nonideality on some physical characteristics in a real fermion gas.

As a result, physical potential and motivation for employing the present deformed quasifermion gas model containing the fermionic Fibonacci oscillators into different areas of research have been enhanced by the obtained results on their applications to thermal and electronic properties of some materials at low temperatures. The developed model here appears to dispose of some of the difficulties involved in earlier models such as the free-electron Fermi theory. Our model could effectively be used to approximate some nonlinear behavior in interacting gas of composite fermions encountered in many interdisciplineary areas of research such as carbon nanotubes, high temperature superconductors, and quantum information based materials.

**Acknowledgments**

We thank the referees for their helpful remarks. This work is supported by the Scientific and Technological Research Council of Turkey (TUBITAK) under Project No. 113F226. We also thank Ali Olkun for his help in preparing the three-dimensional plots.

**References**

[1] Arik M and Coon D D 1976 *J. Math. Phys.* **17** 524

[2] Biedenharn L C 1989 *J. Phys.* A*: Math. Gen.* **22** L873

[3] Macfarlane A J 1989 *J. Phys.* A*: Math. Gen.* **22** 4581

[4] Ng Y J 1990 *J. Phys.* A*: Math. Gen.* **23** 1023

[5] Beckers J and Debergh N 1991 *J. Phys.* A*: Math. Gen.* **24** L1277

[6] Chaichian M, Ellinas D and Kulish P 1990 *Phys. Rev. Lett.* **65** 980

[7] Parthasarathy R and Viswanathan K S 1991 *J. Phys.* A*: Math. Gen.* **24** 613

[8] Chakrabarti R and Jagannathan R 1991 *J. Phys.* A*: Math. Gen.* **24** L711

[9] Arik M, Demircan E, Turgut T, Ekinci L and Mungan M 1992 *Z. Phys.* C **55** 89

[10] Bonatsos D and Daskoloyannis C 1999 *Prog. Part. Nucl. Phys.* **43** 537

[11] Avancini S S and Krein G 1995 *J. Phys.* A*: Math. Gen.* **28** 685

[12] Scarfone A M and Narayana Swamy P 2008 *J. Phys.* A*: Math. Theor.* **41** 275211

[13] Scarfone A M and Narayana Swamy P 2009 *J. Stat. Mech.: Theor. Exp.* P02055

[14] Gavrilik A M and Mishchenko Yu A 2013 *Ukr. J. Phys.* **58** 1171

[15] Gavrilik A M and Mishchenko Yu A 2014 *Phys. Rev.* E **90** 052147

[16] Gavrilik A M and Mishchenko Yu A 2015 *Nucl. Phys.* B **891** 466

[17] Gavrilik A M and Mishchenko Yu A 2012 *Phys. Lett.* A **376** 1596

[18] Tichy M C, Bouvrie P A and Molmer K 2012 *Phys. Rev.* A **86** 042317

[19] Martin-Delgado M A 1991 *J. Phys.* A*: Math. Gen.* **24** L1285

[20] Neskovic P V and Urosevic B V 1992 *Int. J. Mod. Phys.* A **07**, 3379 (1992).

[21] Lee C R and Yu J P 1990 *Phys. Lett.* A **150** 63

   Lee C R and Yu J P 1992 *Phys. Lett.* A **164** 164

[22] Ge M L and Su G 1991 *J. Phys.* A*: Math. Gen.* **24** L721

[23] Song H S, Ding S X and An I 1993 *J. Phys.* A*: Math. Gen.* **26** 5197

[24] Chaichian M, Felipe R G and Montonen C 1993 *J. Phys.* A*: Math. Gen.* **26** 4017

[25] Rego-Monteiro M, Roditi I and Rodrigues L M C S 1994 *Phys. Lett.* A **188** 11

[26] Kaniadakis G, Lavagno A and Quarati P 1997 *Phys. Lett.* A **227** 227

[27] Rego-Monteiro M, Rodrigues L M C S and Wulck S 1996 *Phys. Rev. Lett.* **76** 1098

[28] Altherr T and Grandou T 1993 *Nucl. Phys.* B **402** 195

[29] Gong R S 1995 *Phys. Lett.* A **199** 81

[30] Daoud M and Kibler M 1995 *Phys. Lett.* A **206** 13

[31] Tsallis C 1994 *Phys. Lett.* A **195** 329

[32] Ubriaco M R 1996 *Phys. Lett.* A **219** 205

[33] Ubriaco M R 1996 *Mod. Phys. Lett.* A **11** 2325

[34] Cai S, Su G and Chen J 2010 *Int. J. Mod. Phys.* B **24** 3323

[35] Cai S, Su G and Chen J 2007 *J. Phys.* A*: Math. Theor.* **40** 11245

[36] Arik M and Kornfilt J 2002 *Phys. Lett.* A **300** 392

[37] Lavagno A and Narayana Swamy P 2002 *Phys. Rev.* E **65** 036101

[38] Lavagno A and Narayana Swamy P 2002 *Physica* A **305** 310

[39] Narayana Swamy P 2006 *Int. J. Mod. Phys.* B **20** 697

[40] Narayana Swamy P 2006 *Eur. Phys. J.* B **50** 291

[41] Lavagno A and Narayana Swamy P 2010 *Found. Phys.* **40** 814

[42] Mirza B and Mohammadzadeh H 2011 *J. Phys. A: Math. Theor.* **44** 475003

[43] Rovenchak A 2014 *Eur. Phys. J.* B **87** 175

[44] Dai W S and Xie M 2013 *Ann. Physics* (NY) **332** 166

[45] Chung W S 1999 *Phys. Lett.* A **259** 437

[46] Hung V V, Phuong D D, Thanh L T K 2011 *Proc. Natl. Conf. Theor. Phys.* **36** 140

[47] Sviratcheva K D, Bahri C, Georgieva A I and Draayer J P 2004 *Phys. Rev. Lett.* **93** 152501

[48] Ngu M V, Vinh N G, Lan N T, Thanh L T K and Viet N A 2016 *J. Phys.:Conf. Ser.* **726** 012017

[49] Greenberg O W 1991 *Phys. Rev.* D **43** 4111

[50] Marinho A A, Brito F A and Chesman C 2016 *Physica* A **443** 324

[51] Marinho A A, Brito F A and Chesman C 2014 *J. Phys.:Conf. Ser.* **568** 012009

[52] Marinho A A, Brito F A and Chesman C 2014 *Physica* A **411** 74

[53] Brito F A and Marinho A A 2011 *Physica* A **390** 2497

[54] Marinho A A, Brito F A and Chesman C 2012 *Physica* A **391** 3424

[55] Tristant D and Brito F A 2014 *Physica* A **407** 276

[56] Dehdashti Sh, Harouni M B, Mirza B and Chen H 2015 *Phys. Rev.* A **91** 022116

[57] Adamska L V and Gavrilik A M 2004 *J. Phys.* A*: Math. Gen.* **37** 4787

[58] Algin A 2010 *Commun. Nonlinear Sci. Numer. Simul.* **15** 1372

[59] Gavrilik A M and Rebesh A P 2012 *Mod. Phys. Lett.* B **26** 1150030

[60] Algin A, Arikan A S and Dil E 2014 *Physica* A **416** 499

[61] Algin A, Arik M and Arikan A S 2002 *Eur. Phys. J.* C **25** 487

[62] Jackson F 1909 *Messenger Math.* **38**, 57

[63] Tuszynski J A, Rubin J L, Meyer J and Kibler M 1993 *Phys. Lett.* A **175** 173

[64] Algin A and Senay M 2016 *Physica* A **447** 232

[65] Tritt T M 2004 *Thermal Conductivity: Theory, Properties and Applications* (New York: Kluwer Academic)

[66] Ashcroft N W and Mermin N D 1987 *Solid State Physics* (Philadelphia: Saunders College Pub.)

[67] Kittel C 2005 *Introduction to Solid State Physics* (New York: Wiley)

[68] Huang K 1987 *Statistical Mechanics* (New York: Wiley)

[69] Greiner W, Neise L and Stöcker H 1994 *Thermodynamics and Statistical Mechanics* (New York: Springer)

[70] Pathria R K and Beale P D 2011 *Statistical Mechanics*, *(3$^{rd}$ ed.)* (London: Butterworth-Heinemann)

[71] Hummel R E 2011 *Electronic Properties of Materials* (Berlin: Springer)

[72] Sinha S K 2005 *Introduction to Statistical Mechanics* (New Delhi: Narosa Publishing House)

[73] Girifalco L A 1973 *Statistical Physics of Materials* (New York: Wiley)

[74] Schwabl F 2006 *Statistical Mechanics*, *(2$^{nd}$ ed.)* (Berlin: Springer)

[75] Grusdt F, Shashi A, Abanin D and Demler E 2014 *Phys. Rev.* A **90** 063610

[76] Kondo J 1964 *Prog. Theor. Phys.* **32** 37

[77] Ballantine K E, Donegan J F and Eastham P R 2016 *Science Advances* **2**, number 4 E1501748

[78] Parthasarathy R and Viswanathan K S 2016 arXiv: hep-th/1605.08524v3

[79] Parthasarathy R and Viswanathan K S 1992 IMSc-Preprint-92/57

[80] Strominger A 1993 *Phys. Rev. Lett.* **71** 3397

[81] Dil E 2015 *Can. J. Phys.* **93** 1274

[82] Dil E and Kolay E 2016 *Advances in High Energy Phys.* Article ID 3973706

**List of the figure captions**

**Fig. 1.** The scaled chemical potential $(\mu(T,p,q)/\varepsilon_F)$ as a function of the scaled temperature $(k_B T/\varepsilon_F)$ for various values of the deformation parameters $p$ and $q < 1$.

**Fig. 2.** The scaled chemical potential $(\mu(T,1,1)/\varepsilon_F)$ for an undeformed fermion gas as a function of the scaled temperature $(k_B T/\varepsilon_F)$ for the case $p = q = 1$.

**Fig. 3.** The $p,q$-deformed Sommerfeld parameter $\gamma^{(p,q)}$ of nickel as a function of the deformation parameters $p$ and $q$ for the case $(p,q) < 1$.

**Fig. 4.** The $p,q$-deformed Sommerfeld parameter $\gamma^{(p,q)}$ of cobalt as a function of the deformation parameters $p$ and $q$ for the case $(p,q) < 1$.

**Fig. 5.** The $p,q$-deformed entropy function $[S^{(p,q)}(T)/k_B N^{(p,q)}(0)]$ as a function of the scaled temperature $(k_B T/\varepsilon_F)$ for various values of the deformation parameters $p$ and $q < 1$.

**Fig. 6.** The entropy function $[S^{(1,1)}(T)/k_B N^{(1,1)}(0)]$ for an undeformed fermion gas as a function of the scaled temperature $(k_B T/\varepsilon_F)$ for the case $p = q = 1$.

**List of the table captions**

**Table 1.** Free-electron theory, experimental and the $p,q$-deformed values of the Sommerfeld parameter for the materials nickel and cobalt, respectively.

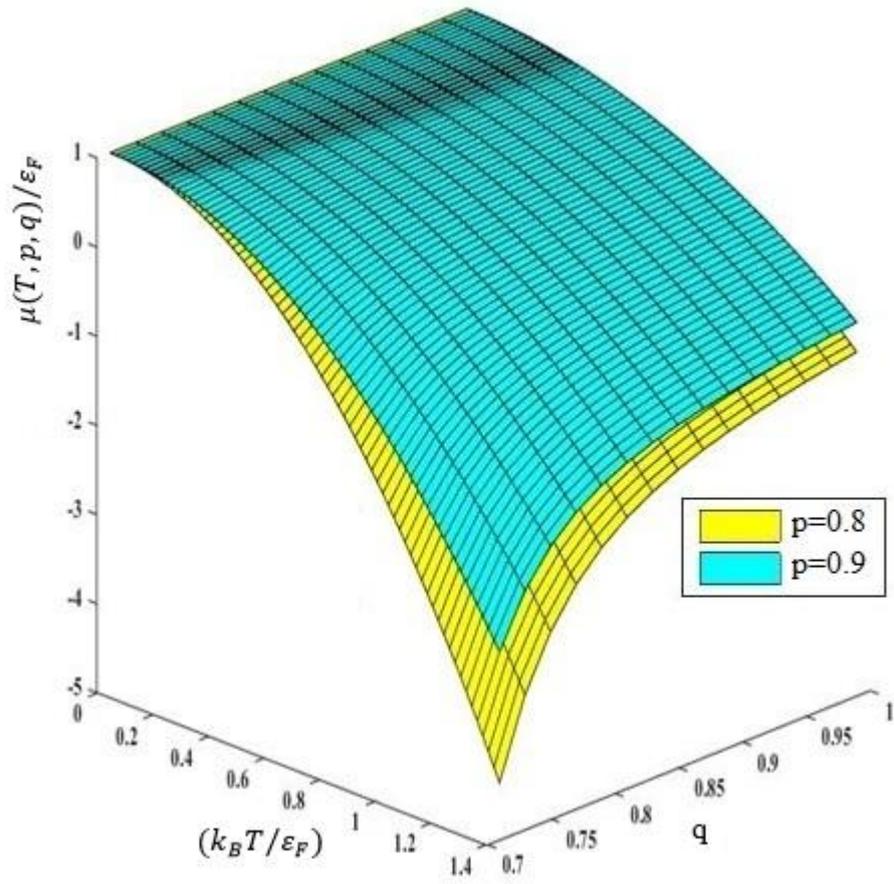

**Figure 1.**

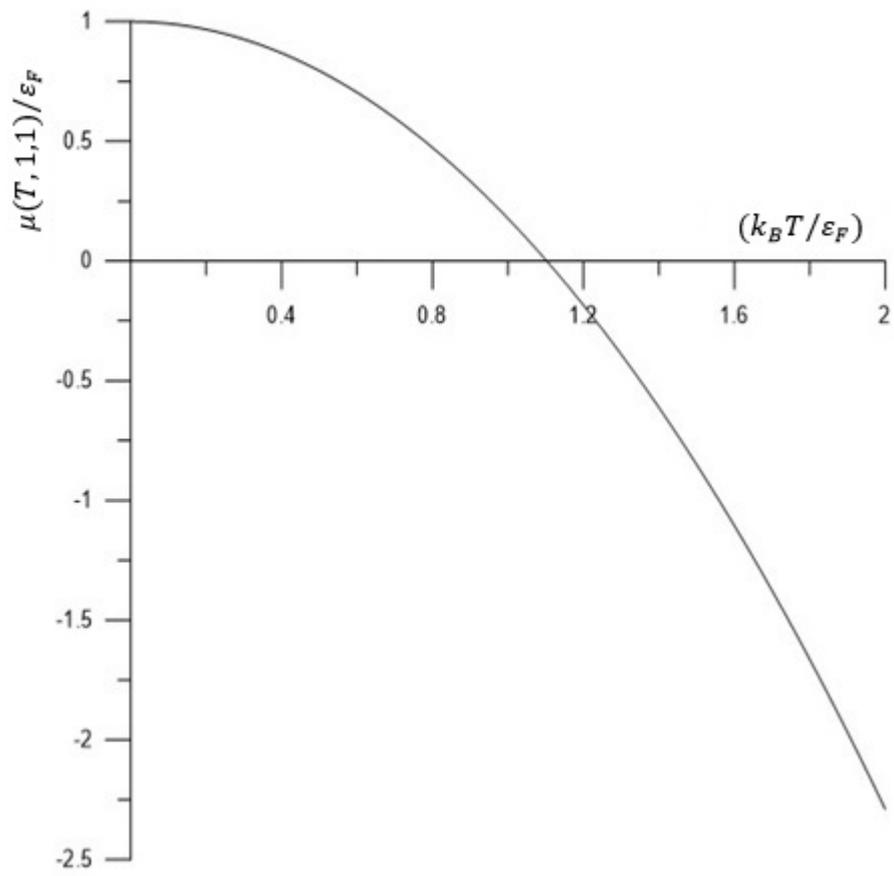

**Figure 2.**

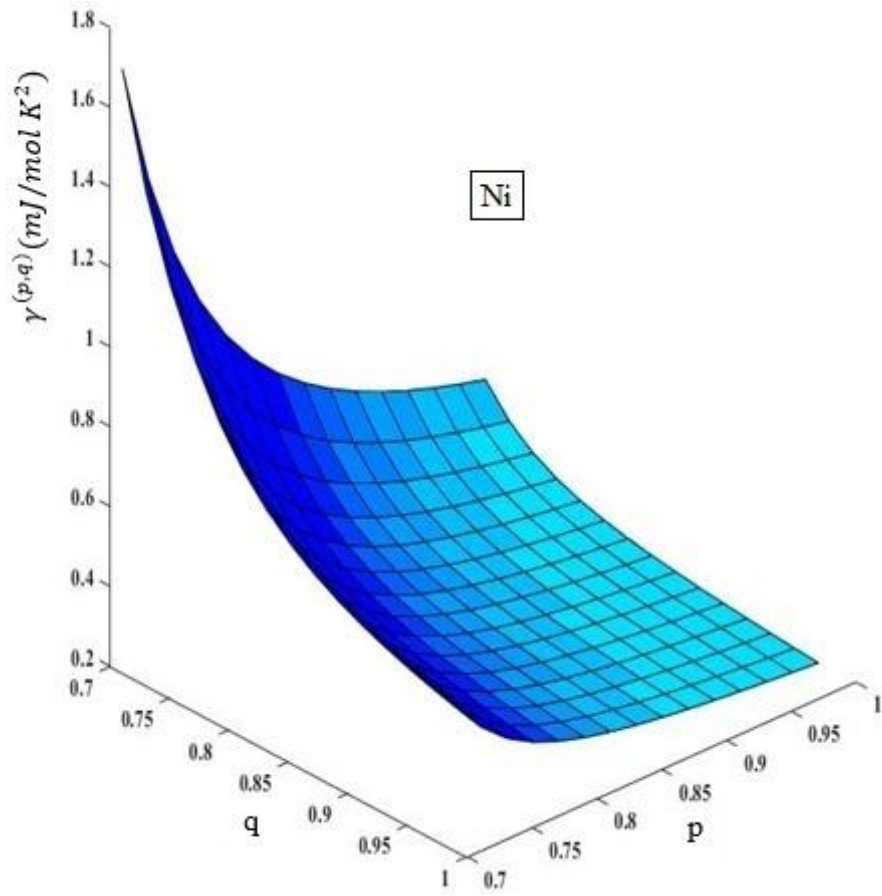

**Figure 3.**

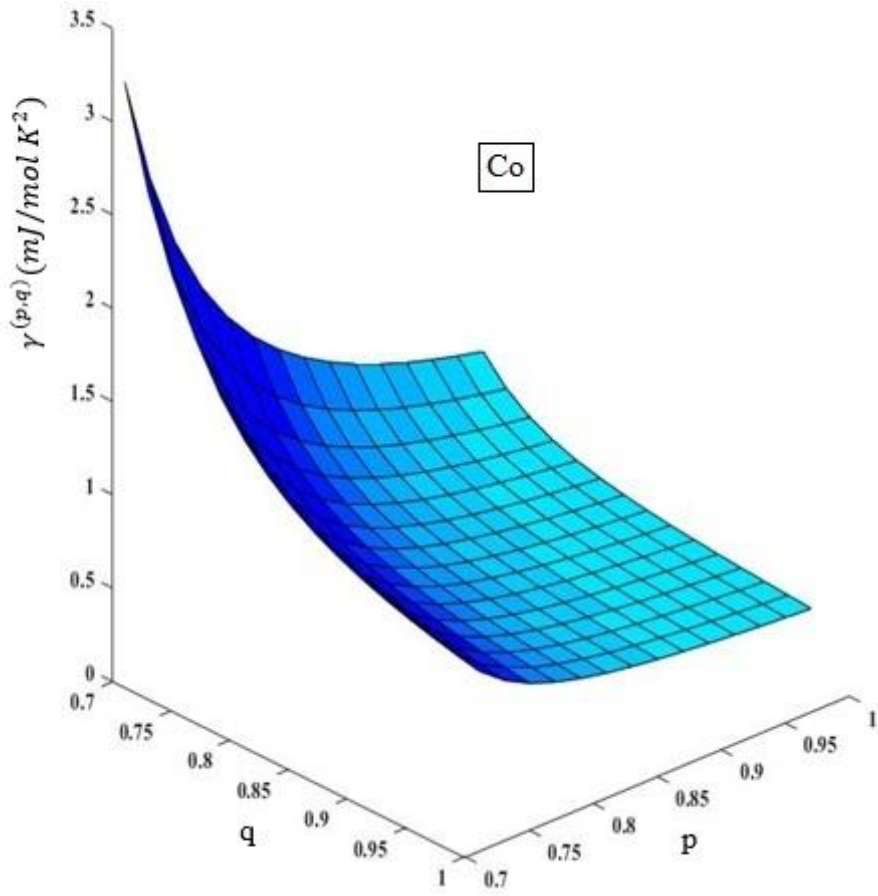

**Figure 4.**

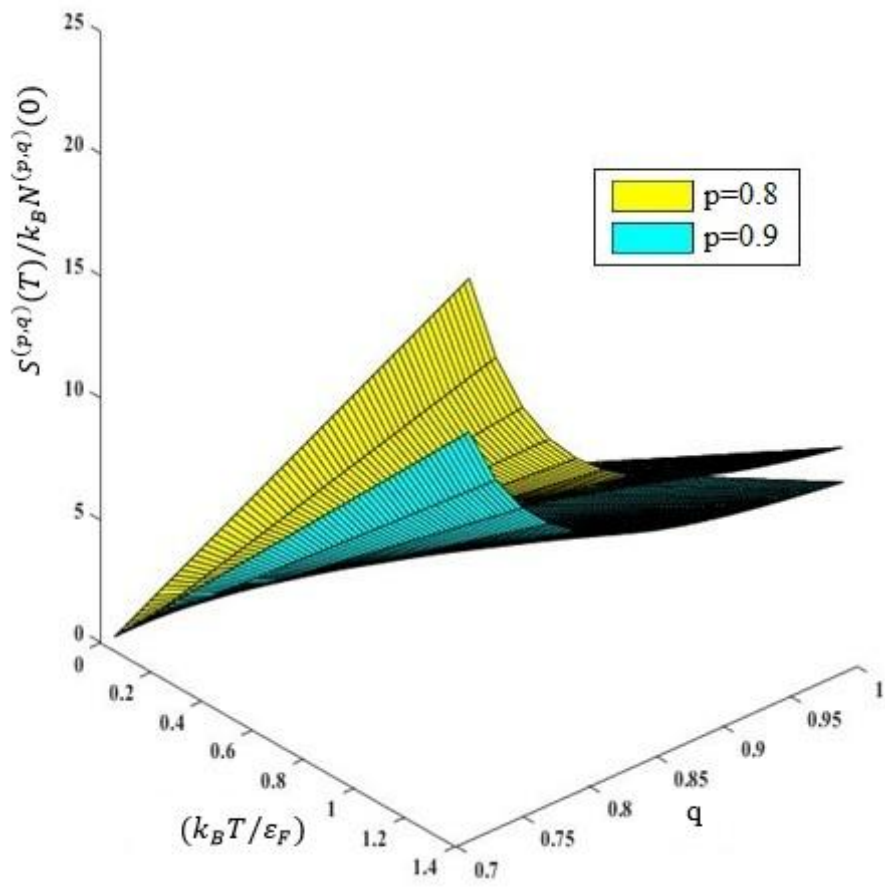

**Figure 5.**

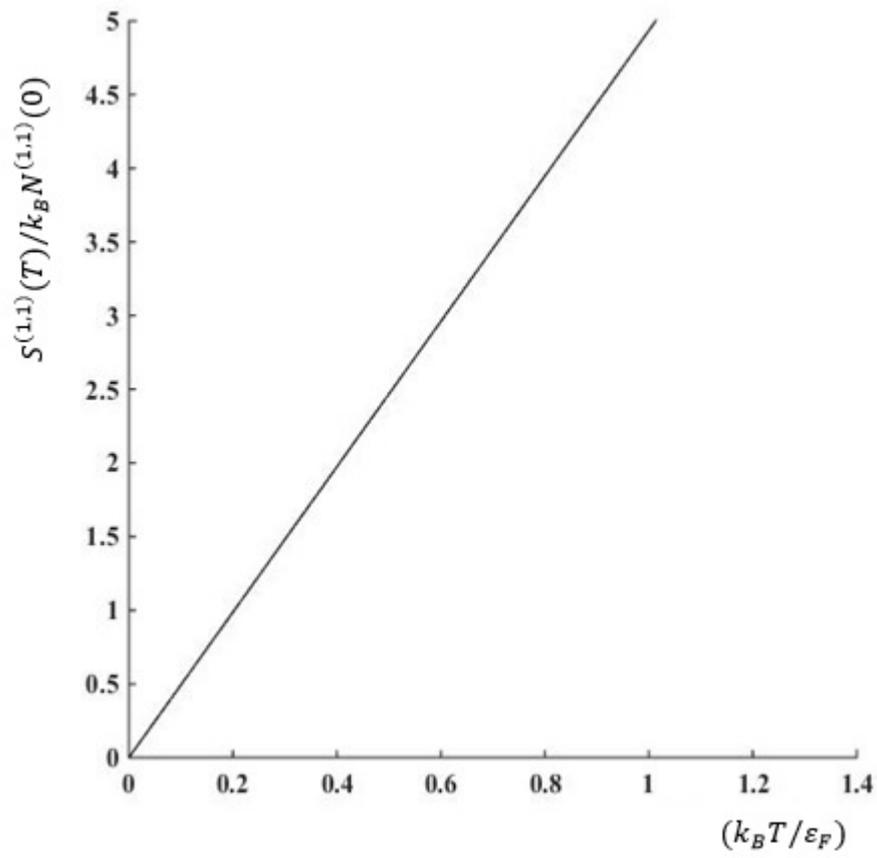

**Figure 6.**

| Material | $\gamma^{(1,1)}$ $(mJ/mol\,K^2)$ [65] | $\gamma^{(\exp)}$ $(mJ/mol\,K^2)$ [65] | $\gamma^{(p,q)}$ $(mJ/mol\,K^2)$ |
|---|---|---|---|
| Ni | 0.61 | 7.02 | 7.02 ($p = 0.8934,\ q = 0.5768$) |
| Co | 0.61 | 4.98 | 4.98 ($p = 0.74,\ q = 0.65$) |

**Table 1.**